\documentclass{mem}
\usepackage{natbib}\usepackage{txfonts}\usepackage{balance}
\usepackage{graphicx}
\usepackage{txfonts}
\usepackage[a4paper]{hyperref}
\idline{79}{3}
\begin{document}
\def\teff{$T\rm_{eff }$}
\def\kms{$\mathrm {km s}^{-1}$}
\def\msol{M$_{\odot}$}

\title{
Lithium abundances in exoplanet-host stars
}


\author{
M. \,Castro\inst{1},
S. \,Vauclair\inst{2},
O. \,Richard\inst{3}
\and N.C. \,Santos\inst{4}
          }

  \offprints{M. Castro}

\institute{
CAAUL - OAL, Tapada da Ajuda, 1349-018 Lisboa, Portugal
\and
LATT - Universit\'e Paul Sabatier, CNRS-UMR 5572, 14 Avenue Edouard Belin, 31400 Toulouse, France
\and
Universit\'e Montpellier II - GRAAL, CNRS-UMR 5024, Place E. Bataillon, 34095 Montpellier, France
\and
CAUP, Rua das Estrelas, 4150-762 Porto, Portugal 
}

\authorrunning{Castro}

\titlerunning{Lithium in exoplanets-host stars}

\abstract{
Exoplanet-host stars (EHS) are known to present surface chemical abundances different from those of stars without any detected planet (NEHS). EHS are, on the average, overmetallic compared to the Sun. The observations also show that, for cool stars, lithium is more depleted in EHS than in NEHS. The overmetallicity of EHS may be studied in the framework of two different scenarii. We have computed main sequence stellar models with various masses, metallicities and accretion rates. The results show different profiles for the lithium destruction according to the scenario. We compare these results to the spectroscopic observations of lithium.
\keywords{Instabilities -- Stars: abundances -- Stars: evolution -- Stars: planetary systems}
}
\maketitle{}

\section{Introduction}

Many studies \citep{santos01,santos03} have shown that exoplanets-host stars (EHS) are on average more metallic by 0.2 dex than stars without any detected planet (NEHS). This involves two scenarii concerning the formation of planetary systems : an original overmetallic protostellar cloud \citep{pinsonneault01, santos01, santos03}, the resulting star is overmetallic from the centre to the surface; or a process of accretion of metallic matter into the star during the formation period of the planetary system \citep{murray01}, the star is overmetallic only in its outer mixed zone.

Precise spectroscopic observations of Li in EHS and NEHS have allowed to take into account new constraints for stellar modelling. Rotation-induced mixing and angular momentum loss seem to be the most efficient processes depleting Li \citep[e.g.][]{stephens97}.\citet{israelian04} compared the Li abundances in a sample including 79 EHS and a comparison sample of 157 NEHS from \citet{chen01}. They give the evidence of an excess of Li-poor EHS (1.0 $< \log N(\rm Li) <$ 1.6) compared to NEHS, concentrated in the effective temperature range 5600-5850 K. 

In this work we compare the evolution of Li (a) in standard models S with a solar metallicity \citep[``old'' abundances from][]{grevesse&noels93}, (b) in initially overmetallic models OM and (c) in models having undergone accretion AC, which simulate the two scenarii of planetary formation. In the next part, we present our models and the results, and in the last part, we give some conclusions.




\section{Modelling} 


We computed the evolution of models with various masses, using the TGEC (Toulouse-Geneva Evolution Code), including microscopic diffusion and rotation-induced mixing with the autoregulation by $\mu$-currents \citep{theado&vauclair03}. The parameters of the mixing are choosen to reproduce the solar Li destruction in a solar model at the solar age \citep[Li/Li$_0$ = 1/140 from][]{grevesse&noels93}. 

\subsection{Influence of metallicity}

\begin{figure}[h!]
\begin{center}
\includegraphics[angle=0,height=5.5cm,width=\columnwidth]{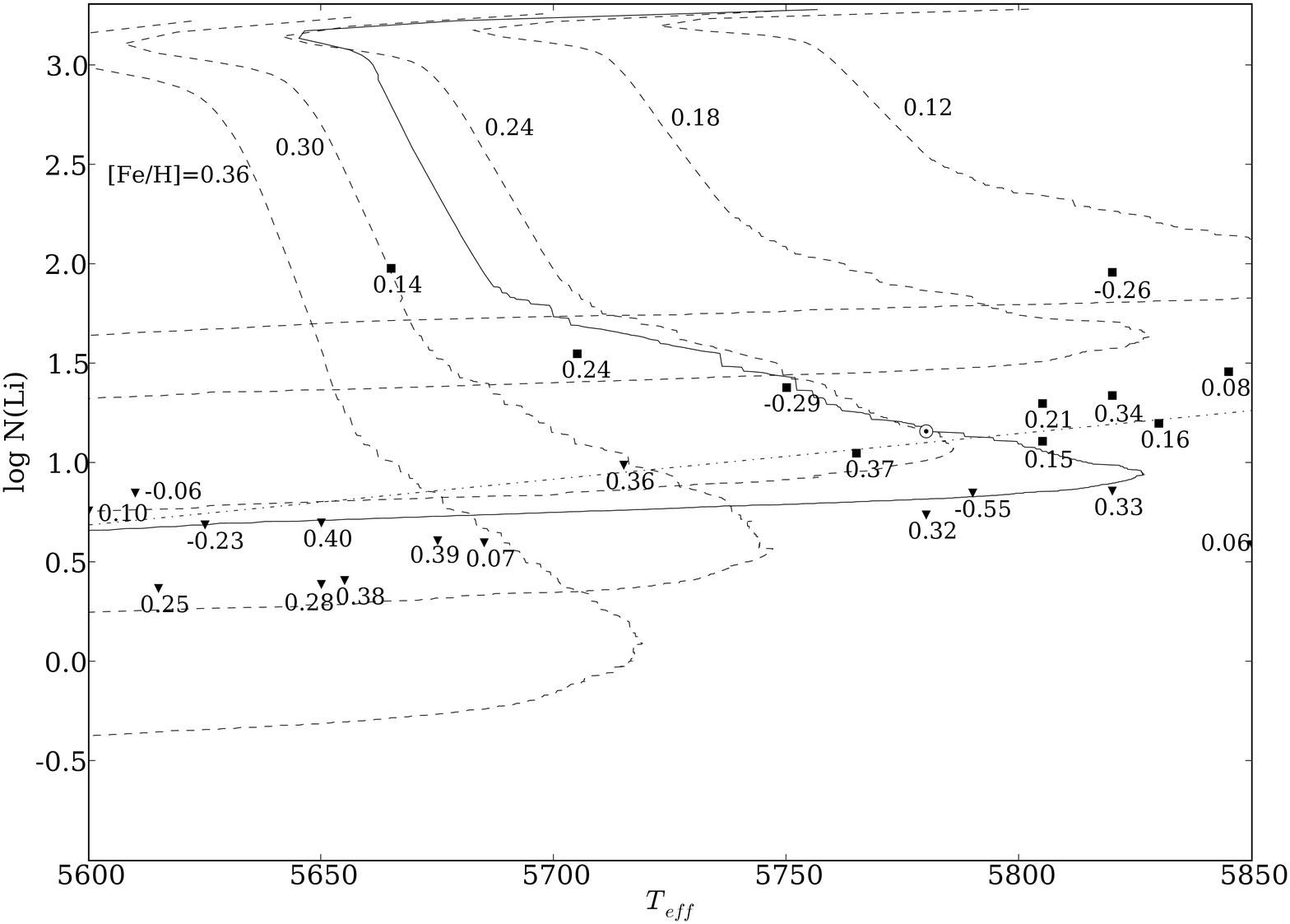}
\end{center}
\caption{\footnotesize
Li destruction along the evolutionnary tracks, as computed by TGEC, as a function of $T_{eff}$, for a model S (solid line) of 1.00 \msol \ and models OM (dashed lines) of 1.05 \msol \ with [Fe/H] = 0.12, 0.18, 0.24, 0.30 and 0.36. The EHS observed by \citet{israelian04} (filled squares and triangles for upper limits), their metallicities and the position of the Sun are indicated. The dot-dashed line is the linear regression of $\log N(\rm Li)$ for the observed stars.}
\label{fig:evolli-metal-sm}
\end{figure}

\begin{figure}[h!]
\begin{center}
\includegraphics[angle=0,height=5.5cm,width=\columnwidth]{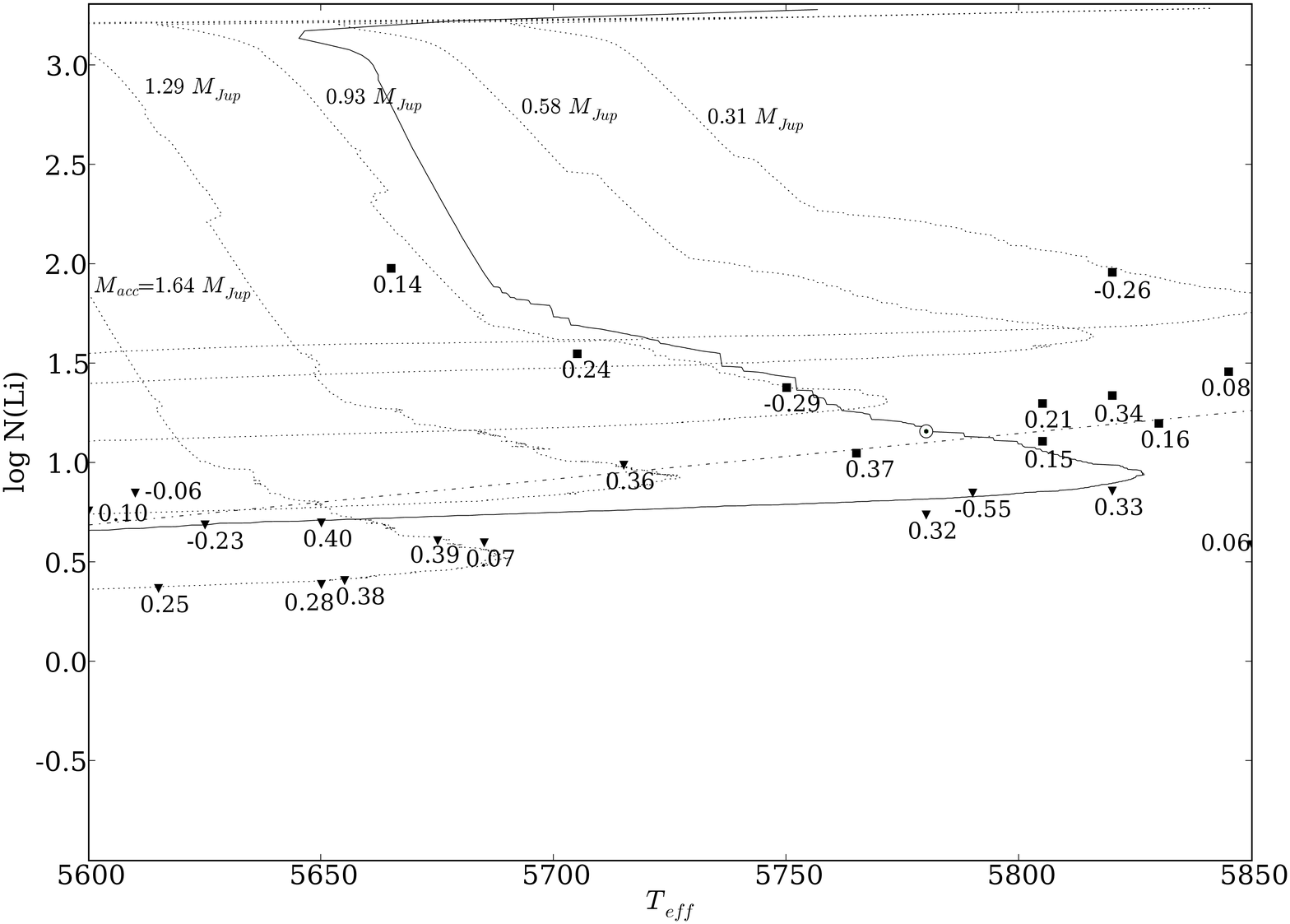}
\end{center}
\caption{\footnotesize
Same as Figure \ref{fig:evolli-metal-sm}, but for a model S (solid line) of 1.00 \msol \ and models AC (dotted lines) of 1.03 \msol with M$_{\rm acc}$ = 0.31, 0.58, 0.93, 1.29 and 1.64 M$_{\rm Jup}$ of metals.}
\label{fig:evolli-metal-ac}
\end{figure}

Figures \ref{fig:evolli-metal-sm} and \ref{fig:evolli-metal-ac} present the Li destruction along the evolutionary tracks for a sample of our computed models, as a function of the effective temperature. We compare the effect of metallicity for models OM of 1.05 \msol \ (Figure \ref{fig:evolli-metal-sm}) and models AC of 1.03 \msol \ (Figure \ref{fig:evolli-metal-ac}). The position of the EHS studied in \citet{israelian04} are also plotted. For these stars (linear regression curve), the Li depletion clearly increases for smaller effective temperatures.

The higher is the surface metallicity, the more is the Li destroyed during the evolution, and the larger is the difference between models OM and AC. The positions of the observed stars in this graph show that those which are cooler than the Sun can be accounted for by models either OM or AC. On the other hand, the stars hotter than the Sun cannot be accounted for in this framework.  

\subsection{Influence of mass}

\begin{figure}[h!]
\begin{center}
\includegraphics[angle=0,height=5.5cm,width=\columnwidth]{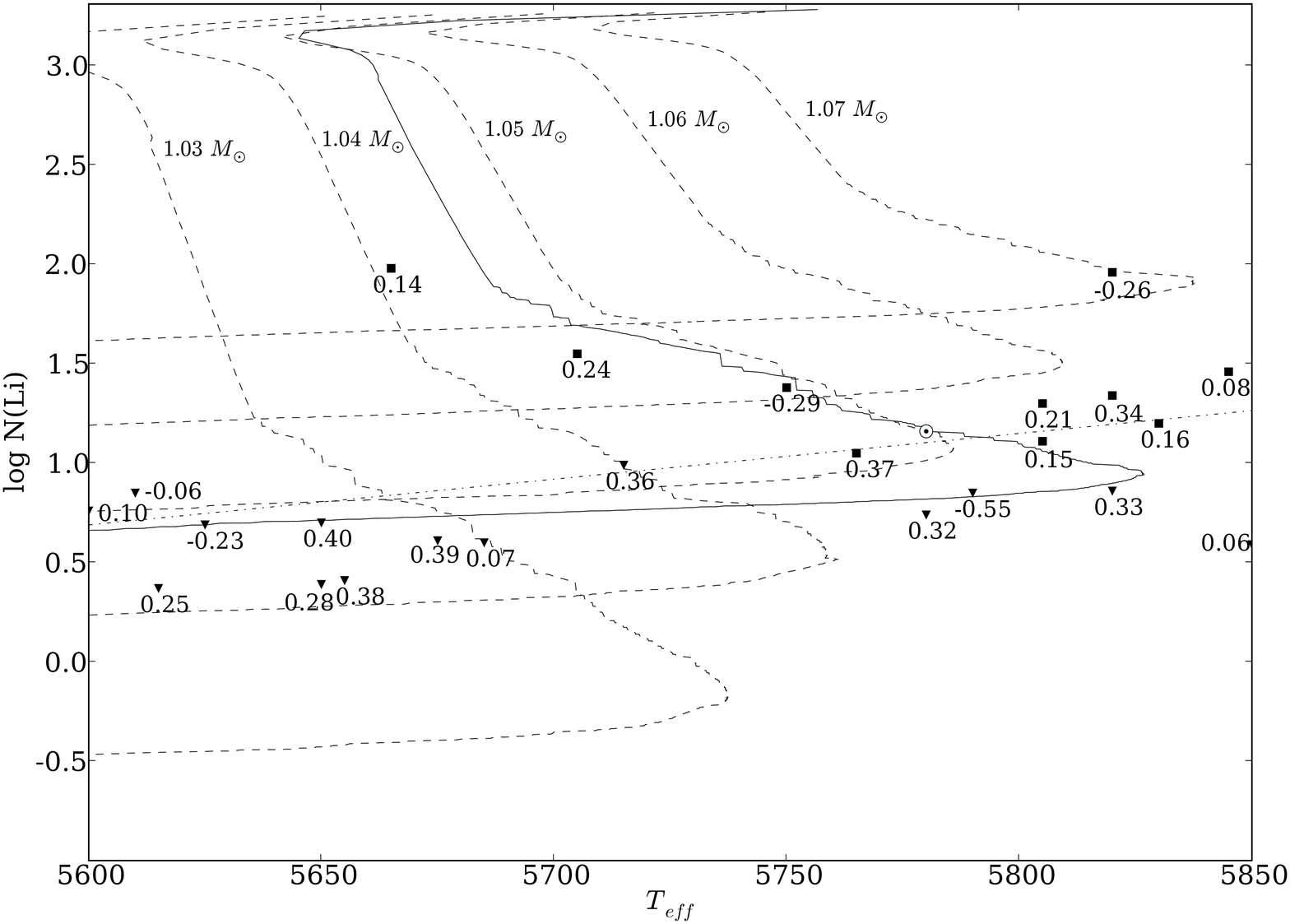}
\end{center}
\caption{\footnotesize
Same as Figure \ref{fig:evolli-metal-sm}, with models of masses 1.03, 1.04, 1.05, 1.06, 1.07 \msol \ and [Fe/H] = 0.24.}
\label{fig:evolli-mass-sm}
\end{figure}

\begin{figure}[h!]
\begin{center}
\includegraphics[angle=0,height=5.5cm,width=\columnwidth]{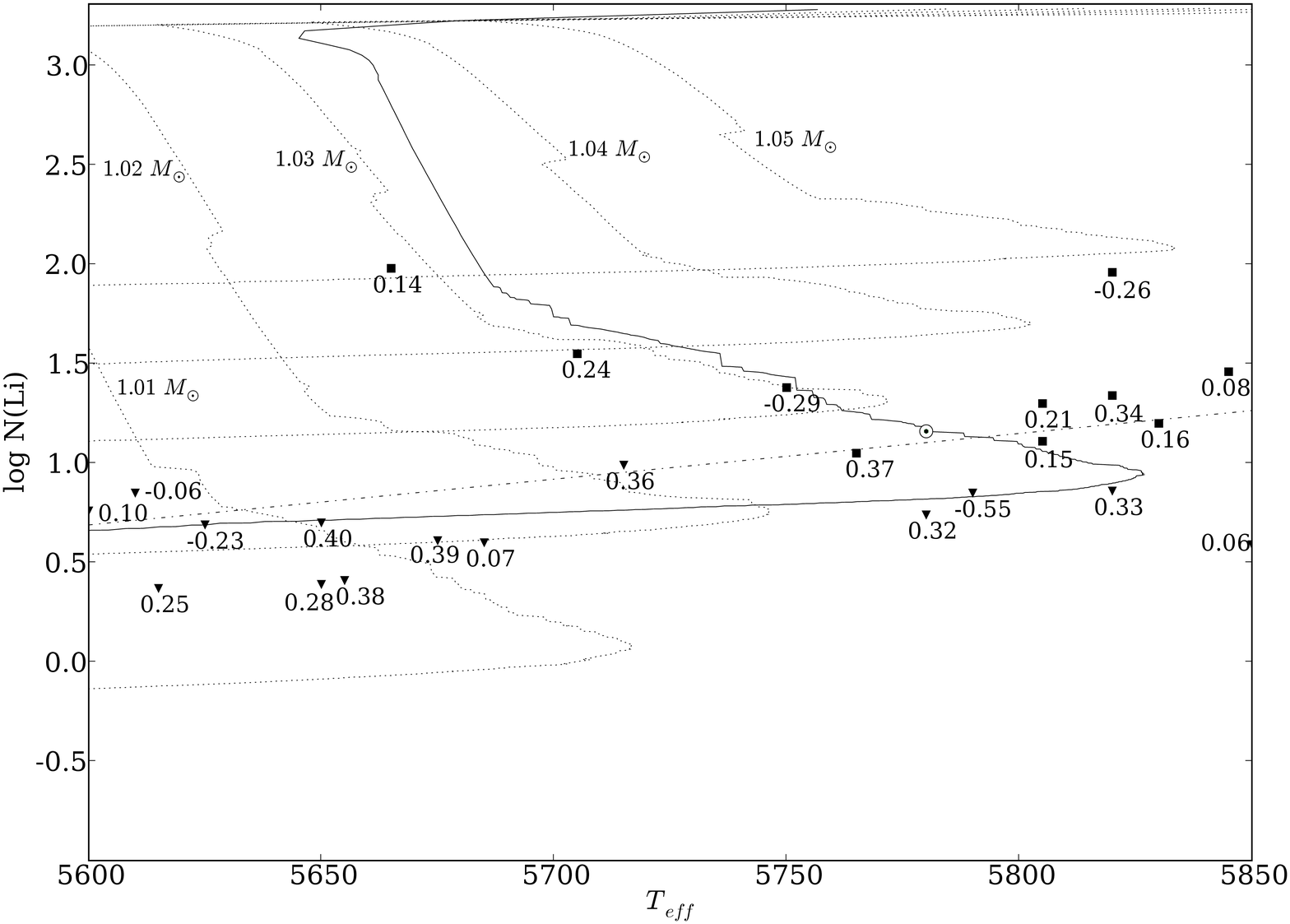}
\end{center}
\caption{\footnotesize
Same as Figure \ref{fig:evolli-metal-ac}, with models of masses 1.01, 1.02, 1.03, 1.04, 1.05 \msol \ and M$_{\rm acc}$ = 0.93 M$_{\rm Jup}$ of metals.}
\label{fig:evolli-mass-ac}
\end{figure}

\begin{figure}[h!]
\begin{center}
\includegraphics[angle=0,height=5.5cm,width=\columnwidth]{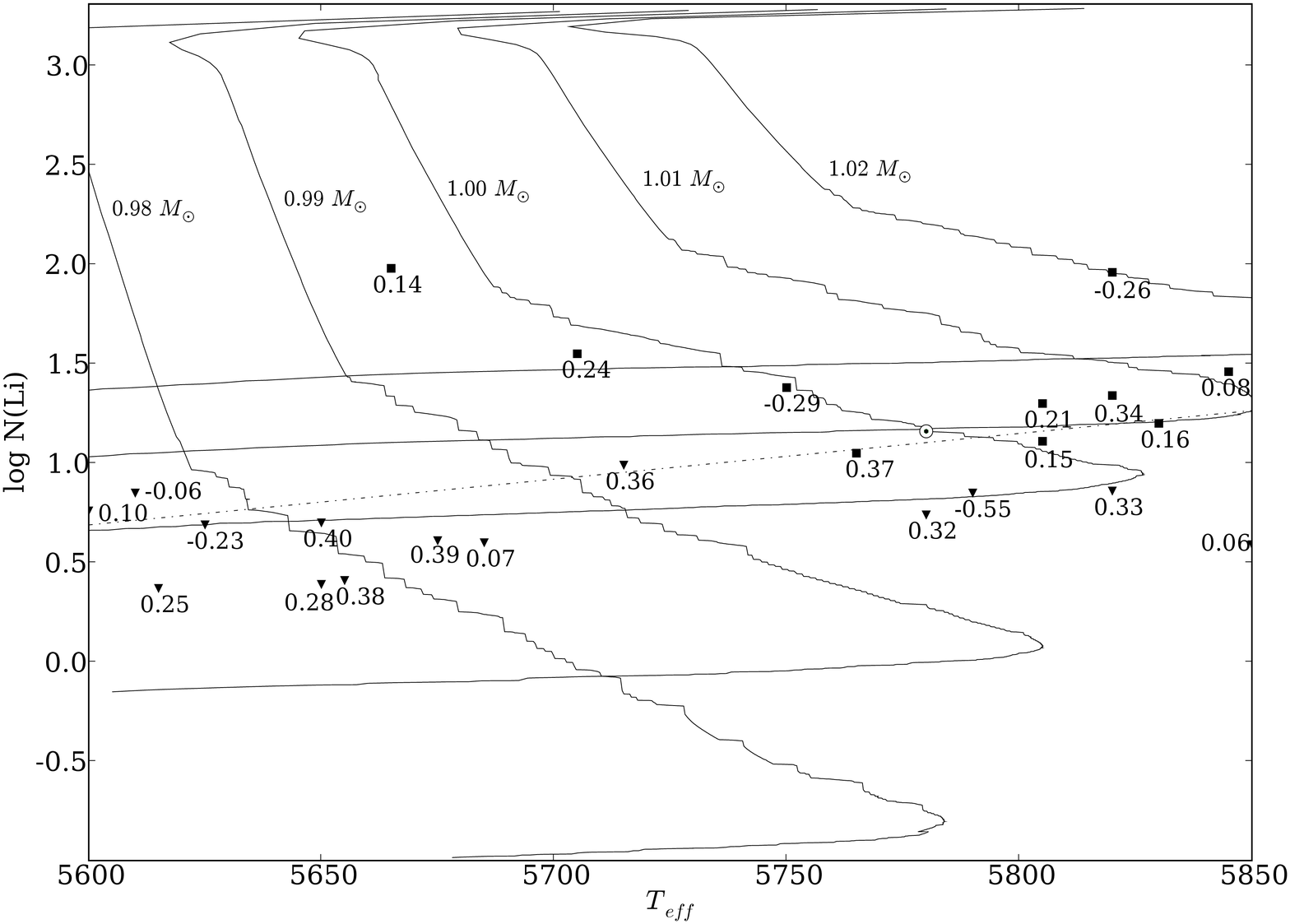}
\end{center}
\caption{\footnotesize
Same as Figures \ref{fig:evolli-metal-sm} to \ref{fig:evolli-mass-ac}, but for models S of masses 0.98, 0.99, 1.00, 1.01 and 1.02 \msol.}
\label{fig:evolli-stand}
\end{figure}

Li destruction in models OM (Figure \ref{fig:evolli-mass-sm}) and models AC (Figure \ref{fig:evolli-mass-ac}) with the same metallicity ([Fe/H] = 0.24, corresponding to an accreted mass of M$_{\rm acc}$ = 0.93 M$_{\rm Jup}$ of metals) and different masses, are presented as a function of the effective temperature. Figure \ref{fig:evolli-stand} presents the destruction of Li in models S.

The lower is the mass of the model, the larger is the Li destruction during evolution. The difference between the models OM and AC does not depend on the mass. As in Figures \ref{fig:evolli-metal-sm} and \ref{fig:evolli-metal-ac}, the hottest observed stars cannot be accounted for by models either OM or AC. We can see in Figure \ref{fig:evolli-stand} that the Li destruction in these stars could be accounted for by models S with masses between 1.00 and 1.02 \msol, but this is inconsistent with their observed metallicities.

\section{Conclusion}

While the Li depletion in stars cooler than the Sun, as observed by \citet{israelian04} could be accounted for in the framework of models either OM or AC, it is not possible for hotter stars. The Li depletion observed in EHS hotter than the Sun cannot be directly related to their overmetallicities. This result is in favour of the suggestion by \citet{israelian04} and \citet{chen&zhao06} that the strong Li depletion observed in these stars could be related to the planet migration mechanism: an angular momentum transfer could lead to shear-flow instabilities below the convective zone, which would induce an extra-mixing process at the beginning of the star's evolution on the main-sequence.


\bibliographystyle{aa}

\end{document}